
\input phyzzx

\magnification=1200
\overfullrule=0pt
\nopagenumbers
\baselineskip = 12pt
\vsize 25 truecm
\hsize 17 truecm
\hoffset -0.5 truecm
\voffset -0.3 truecm
\def\tvi{\vrule height 12pt depth 6pt width
 0pt}
\def\tv{\tvi\vrule}
\def \cc #1 {\kern .7em \hfill #1 \hfill \kern .7em}

\vbox to 4 truecm {}
\centerline{\bf SEARCHING FOR AN INTERMEDIATE-MASS HIGGS}
 \centerline{\bf AND NEW PHYSICS IN TWO-PHOTON COHERENT}
\centerline{\bf PROCESSES AT THE LHC.} \par
\vskip 5 mm
\centerline{\bf Elena Papageorgiu} \par
\centerline{Laboratoire de Physique
Th\'eorique et Hautes Energies\footnote{*}
{Talk pres. at the XXXth Rencontres de Moriond
on QCD and High Energy Hadronic Interactions,
March 19-26, Les Arcs, 1995.}}
\par \centerline{Universit\'e de Paris XI,
b\^atiment 211, 91405
Orsay Cedex, France} \par

\vbox to 6 truecm {}
{\bf Abstract} \par
We reexamine the prospects for searching for a neutral Higgs boson
in the intermediate-mass range and physics beyond the standard model, using the
proton- and ion-beam facilities of the LHC
to study coherent two-photon processes.
Considering realistic design luminosities
for the different ion beams, we find that beams of
light-to-medium size ions like calcium will give the
highest production rates.
With a suitable trigger and assuming a b-quark identification
efficiency of $30\%$ and a $b{\bar b}$ mass resolution of 10 GeV
one could expect to see a Higgs
signal in the $b{\bar b}$ channel with
a 3-4 $\sigma$ statistical significance in
the first phase of operation of the LHC with beams of calcium.
The discovery potential for light supersymmetric particles is as promissing.
The study of such very ``quiet'' final state topologies in heavy
ion collisions could even lead to the discovery of a new phase of QED.

\vskip 5 mm
\noindent {\bf ORSAY-Preprint 95-46}
\vskip 2 mm
\noindent June 1995
\vfill \supereject
\baselineskip=18 pt

\noindent
{\bf (I) An intermediate-mass Higgs in coherent NN and pp collisions at the
LHC.}
\par
\vskip 3mm
The search for the Higgs particle(s) of the standard model (or its
supersymmetric version) beyond the discovery potential of LEP2
has become one of the major objectives of the LHC.
The intermediate-mass range $M_H \simeq 80 - 130$ GeV is particularly
challenging for both theory and experiment. Here, simultaneous searches in
different channels of hard pp collisions and a minimum of three years of
operation at the  initial luminosity $L_p \simeq 10^{33} cm^{-2} s^{-1}$ will
be needed in order to
establish a 3-4 $\sigma$ signal, or, one will have to wait for the design
luminosity
$L_p^d \simeq 10^{34} cm^{-2} s^{-1}$ by the year $2008$ [1]. \par

A different kind of Higgs- and SUSY-particle production,
which would make use of the ion beam facilities at the LHC,
was proposed some time ago [2], namely, the two-photon production in the
coherent electromagnetic field of nucleus nucleus collisions in
which the nuclei $N_{i=1,2}$ would ``ideally'' remain intact~:

$$
N_1N_2 \to N_1N_2 + X \qquad  X=H,{\tilde \ell}^+ {\tilde \ell}^-,
\tilde{\chi}^+ \tilde{\chi}^-, \cdots \,, \eqno[1]
$$

\noindent where ${\tilde \ell}^+ {\tilde \ell}^-$ and
$\tilde{\chi}^+ \tilde{\chi}^-$ refers to a pair of sleptons and
charginos respectively. Triggering exclusively on such events
would imply cross sections which scale with the nuclear charges as
$Z_1^2Z_2^2$ and a much cleaner (purely photonic) environement for
particle searches than the usual hadronic background of pp collisions. The
``coherency'' of the
collision can be guaranteed by imposing cuts on the impact
parameter space of the two nuclei (Fig~1)~:

$$
b_i > R_i \qquad {\rm and} \qquad |b^{\to}_1 - b^{\to}_2| > R_1 + R_2 \,,
\eqno[
2]
$$

\noindent {\it i.e}, treating the nuclei as two nonoverlapping, ``hard'' and
``opaque''
discs with radii $R_i\simeq r_0 A_i^{1/3}$ ($r_0 \simeq 1.2$ fm) [3,4]. One
should notice
 that when nuclei or nucleons overlap, elastic scattering is
partly the ``shadow'' of inelastic processes.
Therefore the presence of the nuclear elastic form factor is not sufficient
to exclude strong interaction processes taking place
after the photons have been emitted from the nuclei. The constraints of eq.(2)
imply almost real photons, $q_i^2 < {1\over R_i^2}$ and the
factorization of the cross section in the underlying subprocess:
$\sigma(\gamma\gamma \to X)$ times the
two-photon flux [3]

$$\eqalignno{
{dL_{\gamma \gamma} \over dW^2} &= \int \int f_1(\omega_1, b_1) \
f_2(\omega_2 , b_2) \ \Theta (B - R_1 - R_2) \cr
&\hskip 1.7 truecm \times \delta (W^2 - 4 \omega_1 \omega_2) \ d \omega_1 \ d
\omega_2 \  d^2b_1
\  d^2b_2 &[3]  \cr  }$$ \par

\noindent where $f_i (\omega_i, b_i)$ is the number of photons with energy
$\omega_i$ at a fixed
impact and $W^2 = 4 \omega_1 \omega_2$ is the invariant mass squared.
\par

For a Higgs in the intermediate-mass range which would predominantly decay into
a $b{\bar b}$ pair the signal-to-background ratio is sizably more
favourable in coherent collisions than in hard-scattering
NN and pp collisions~:

$$\eqalign{
&(S/B)_{COH} \equiv {\gamma\gamma \to H \to b{\bar b} \over
\gamma\gamma \to b{\bar b}} \simeq 10^{-1} \cr
&(S/B)_{HS} \equiv {gg \to H \to b{\bar b}\over gg \to b{\bar b}} \simeq
10^{-3}-1
0^{-4}\,, \cr
}  \eqno[4]
$$

\noindent assuming a $b{\bar b}$ resolution of ${\cal  R}\simeq 10$ GeV.
This is mainly due to the difference in electric charge between the
bottom and the top quark, the latter being the dominant contribution
in the  $\gamma\gamma \to H$ loop.
Imposing cuts on the rapidity and the transverse momentum of the b jets can
improve significantly these ratios [5], {\it e.g.}, for a $p_T \geq 0.4 M_H$:

$$
\eqalign{
(S/B)_{COH} &\simeq {1\over 5}
\qquad {\rm 100 GeV} < {\rm M_H} < {\rm 120 GeV} \cr
&\simeq {1\over 3} \qquad {\rm 120 GeV} < {\rm M_H} < {\rm 150 GeV} \cr
} \eqno[5]$$

To take advantage of the very favourable signal-to-background ratio in coherent
processes one would need a powerful
veto trigger on spectator jets, coming
from diffractive dissociation and/or nucleon fragmentation of the beam,
which will go in the very forward (backward) direction and will have a typical
transverse momentum of a few GeV.
If instead of ion beams one would use proton beams, tagging could be possible.
An extension of the ALICE detector with
forward spectrometers for studying among other
physics genuine two-photon processes in pp and NN collisions is currently under
discussion [6]. The
production of muon pairs would be best suited for measuring the two-photon flux
in heavy-ion
collisions already at the present energy of the SPS, where one could also study
the spectroscopy of low-energy
resonances like the $\eta '$. \par

The crucial point of course is whether the production rate in coherent
collisions is large enough to see a signal with a statistical
significance of say 3-4 already during the first operational phase of
the LHC, and with which type of beam one would do the best job.
When this mechanism was originally proposed it was thought that the use
of heavy ions like $Pb$ would be more advantageous due to the
higher charge. As a matter of fact, in $Pb-Pb$ collisions
the Higgs-boson production is as large as a few tens of picobarns [Fig.~3]
and comparable to the one in hard pp collisions. On the other
hand, a more recent study [7] has shown that the maximum achievable
luminosity $L_b$ is for beams of realy heavy ions several orders of
magnitude lower than for lighter ions or protons (Table 1) due to
large intra beam effects. \par
\baselineskip = 12 pt
$$\vbox{\offinterlineskip \halign{
\tv# &\cc{#} &\tv# &\cc{#} &\tv# &\cc{#} &\tv# &\cc{#} &\tv# \cr
\noalign{\hrule}
& &&{\bf p-p} &&{\bf Ca-Ca} &&{\bf Pb-Pb} & \cr
\noalign{\hrule}
&${\bf L_b [cm^{-2} s^{-1}]}$ &&$10^{33} - 10^{34}$ &&$5 \times 10^{30}$ &&$5
\times 10^{26}$ &
\cr  \noalign{\hrule}
&${\bf E_b [TEV]}$ &&7 &&140 &&574 & \cr
\noalign{\hrule}
&${\bf {\cal W}_0 [GEV]}$ &&$3 \times 10^3$ &&370 &&170 & \cr
\noalign{\hrule}
&${\bf Ev./y}$ &&- (30-70) &&20-50 &&$\ll 1$ & \cr
\noalign{\hrule}
&${\bf S/\sqrt{\bf S + B}}$ &&-- $(6-7)$ &&$5-6$ &&--- & \cr
\noalign{\hrule}
&${\bf (S/\sqrt{\bf S + B})^*}$ &&-- (3-4) &&3-4 &&--- & \cr
\noalign{\hrule}
}}$$
\noindent{\bf Table 1 :} The beam luminosity $L_b$ and the energy per beam
$E_b$ for different type
of collisions at the LHC, from Ref. [7]. The mass range ${\cal W}_0$ which can
be typically
explored via coherent two-photon processes. The number of events per year
for $\gamma \gamma \to H \to b \bar{b}$
 and $M_H \simeq$ 80 - 180 GeV. The statistical significance $S/\sqrt{S + B}$
for
the same process and $M_H \simeq$ 100-130 GeV after 3-4 years of running. The
same as above
but assuming 30 $\%$ $b$-detection efficiency.  \par \vskip 3 mm

\baselineskip=18 pt
Another reason why lighter ions or even protons may be preferable is the
following. The typical
mass range which can be explored via two-photon coherent pocesses in hadron
collisions is
limited to :

$${\cal W}_0 = {2 \gamma \over \sqrt{R_1R_2}} \,, \eqno[6] $$

\noindent simply because the number of equivalent photons
in the electromagnetic pulse created by relativistic nuclei is
decreasing rapidly for $\omega > \gamma/R$. At the LHC where the maximum
energy of a proton beam will be $E_p = 7$ TeV, the maximum energy of an
ion beam will be $E_{ion} = E_p Z$ and
$\gamma \simeq \, 7.5 \, {Z\over A} \,\,[{{\rm TeV \over n}}]$.
Typically the mass range which can be explored with
different type of beams is shown in Table 1.
For the same reason the two-photon flux is decreasing rapidly as
the nuclear size increases for invariant masses close to ${\cal W}_0$,

$$
{dL_{\gamma \gamma} \over dW^2} = {{\rm const.}\over W^2}
\times {\cal F} \left ( {{\cal W}_0 \over W} \right ) \,. \eqno[7] $$ \par

\noindent A comparison of the two-photon luminosity function
$dL/dW = L_b \times {dL_{\gamma
\gamma} \over dW}$ for heavy ions ($Pb$-$Pb$), medium-size ions ($Ca$-$Ca$) and
protons ($p$-$p$)
shows that beams of light-to-medium-size nuclei would be the better choice. The
upper and lower $pp$
curves correspond to a cut-off of $R \simeq 0.2$ fm, which is the proton radius
as determined from
elastic electron scattering, and, a cut-off $R \simeq 1$ fm at the outer edge
of the very extended
proton surface.
They show that applying less stringent cuts one would gain a factor ten
more in $\gamma\gamma$ luminosity, at the expense of some hadronic
background. Since a veto detector on spectator jets cannot
exclude partial or full dissociation of the nuclei, very soft
hadronic processes, and final state interactions, the precise value
of the theoretical cut-off can be chosen only in connection with the
performance of the trigger for which a simulation would be needed [8].\par

The total cross section for producing an intermediate-mass
Higgs in coherent $Pb-Pb$, $Ca-Ca$ and $p-p$ csions is shown in
Fig. 3 and the corresponding event rates for one year of full running
are given in Table 1. The event rate in $Ca-Ca$ collisions is comparable
to the one in $p-p$ collisions at the higher design luminosity.
For such collisions the statistical significance that can be
achieved after one year will be $S/\sqrt{S+B}\simeq 2$, but, assuming
realistic values for the b-detection efficiency of $\epsilon = 30\%$
reduces the signal-to-background ratio of eq.(5) to:

$$\eqalign{
(S/B)_{COH}^{\star} &\simeq {1\over 10}
\qquad {\rm 100 GeV} < {\rm M_H} < {\rm 120 GeV} \cr
&\simeq {1\over 5} \qquad {\rm 120 GeV} < {\rm M_H} < {\rm 150 GeV}\,.
\cr} \eqno[8] $$

\noindent With this one could reach the same statistical significance
as in the hard-scattering pp channels, namely
$(S/\sqrt{S+B})^{\star}\simeq 3-4$
within the operational phase I of the LHC, {\it i.e}, within three
to four years.\par
\vskip 5 mm

\noindent
{\bf (II) Two-photon production of non-strongly interacting SUSY
particles at the LHC.}\par
\vskip 3mm
In contrast to coloured supersymmetric particles (squarks and gluinos)
which will be produced copiously at the LHC, the search for the
non-strongly interacting
supersymmetric particles (sleptons and charginos)
in the decay cascades of the former will be particularly
challenging in hard pp collisions. Searches in the much cleaner
environement of coherent hadron collisions, and given the fact that
their coupling to photons is well known
-charginos couple as fermions ($f^+f^-$) and sleptons as scalars
($S^+S^-$)-, could be more promissing. The corresponding cross sections
for Ca-Ca collisions are shown in Fig. 4. For charginos with a mass
$100 GeV\leq M_{\tilde{\chi}}\leq 150 GeV$ one should expect at least
 ${\bf (50 - 5) Ev./y}$ while the sleptons rate in the mass range
$100 GeV\leq M_{\tilde \ell}\leq 120 GeV$ will be smaller,
${\bf (15 - 5) Ev./y}$. According to Fig. 2 the same rates are expected
also for coherent pp collisions for an integrated luminosity of $100
fb^{-1}$.

\noindent  Our predictions, for which the impact parameter cut-off
was used in order to exclude all strong absorbtion processes, lie below
the predictions in ref.[9] where the elastic
formfactor of the proton was used. For a comparision
of the different approaches see ref. [3].
The background from $\gamma\gamma \to W^+W^-$
and $\gamma\gamma \to \mu^+\mu^-$, after rapidity and $p_T$ cuts
have been applied, can be successfully suppressed with respect to
the signal by requiring
the transverse momenta of the $W$'s or better of the $\mu$'s not to
balance ({\it e.g.} within 10 GeV)  -as expected from the coherency
condition of eq.(2)- due
to the escaping invisible LSP's (the lightest stable supersymmetric
particle) [9]. In this way and assuming an error of $\pm 5$ GeV the search
for a light chargino (slepton) in coherent collisions of protons and
light-to-medium size nuclei should become a feasible task. \par

\noindent
{\bf (III) Conclusions}\par
Using the proton and ion beam facilities at the LHC to study
very ``quiet'' final state topologies, so-called coherent processes,
could lead to some nice surprises, ranging from the discovery
of a 100-130 GeV mass Higgs boson and light supersymmetric particles, to
the yet more exotic possibility of a new (nonperturbative,
chiral) phase of QED reserved to heavy ions only [10].

\baselineskip=12 pt
\vskip 5 mm
\noindent {\bf References}
\vskip 3 mm
\item {[1]} ATLAS Collaboration, CERN-LHCC Report 94-43;
CMS Collaboration, CERN-LHCC Report 94-38. \vskip 2 mm

\item {[2]} E. Papageorgiu, in proc. of the {\it 24th International
Conference on High-Energy Physics}, August 4-10, 1988, Munich,
ed. by R. Kotthaus and J.H. K\"uhn;
E. Papageorgiu, Nucl. Phys. {\bf A498} (1989) 593c;
E.~Papageorgiu, Phys. Rev. {\bf D40} (1989) 92;
M.~Grabiak, B.~M\"uller, W.~Greiner, G.~Soff, and P.~Koch,
J. Phys. {\bf G15} (1989) L25;
E.~Papageorgiu, Max-Planck Report \-MPI-PAE/PTh 68/89.  \vskip 2 mm

\item {[3]} E.~Papageorgiu, Phys.  Lett. {\bf B250} (1990) 155.  \vskip 2 mm

\item {[4]} R.~N. Cahn and J.~D. Jackson, Phys. Rev. {\bf D42} (1990) 3690;
G.~Baur and L.~G.~Ferreira Filho, Nucl. Phys. {\bf A518} (1990) 786;
K. J. Abraham, R. Laterveer, J. Vermaseren and D. Zeppenfeld, Phys.
Lett. {\bf B251} (1990) 186; B. M\"uller and A.~J. Schramm, Phys. Rev. {\bf
D42} (1990) 3699.
\vskip 2 mm

\item {[5]} M.~Drees, J.~Ellis, and D.~Zeppenfeld,
Phys.\ Lett. {\bf B223} (1989) 454;
K. J. Abraham, M. Drees, R. Laterveer, E. Papageorgiu,
A. Sch\"afer, G. Soff, J. Vermaseren and D. Zeppenfeld,
in proceedings of the {\it Large Hadron Collider Workshop},
October 4-9, 1990, Aachen, Vol. II; G.~Altarelli and M.~Traseira, Phys. Lett.
{\bf B245} (1990) 658.\vskip 2 mm

\item {[6]} K. Eggert, Talk pres. at the workshop on Further Physics Topics for
the LHCC, November 29, CERN, 1995, and, priv. communication.\vskip 2 mm

\item {[7]} D. Brandt, K. Eggert, and A. Morsch,
Cern Report \- CERN AT/94-05 (DI), LHC Note 264.
\vskip 2 mm

\item {[8]} E.~Papageorgiu, Phys.  Lett. {\bf B352} (1995) 394.  \vskip 2 mm

\item {[9]} J. Ohnemus, T. Walsh and P. Zerwas, Phys.  Lett. {\bf B328} (1994)
369.  \vskip 2 mm

\item {[10]} E.~Papageorgiu, Phys.  Lett. {\bf B291} (1992) 180;
 E. Papageorgiu, ``Signals for a new phase of QED in elastic scattering of
heavy nuclei'', \-{\bf ORSAY-Preprint 95-47}, {\it to appear in proceedings of
the VIth International
Conference on Elastic and Diffractive Scattering}, Ch\^ateau de Blois, June
20-24, 1995. \vskip 2 mm

\endpage

\noindent
{\bf Figure Captions}

\vskip 2mm
\noindent
{\bf Fig. 1}
Higgs production in coherent nucleus nucleus collisions Feynman graph
(diagram on the left)
and impact parameter plane (diagram on the right).

\noindent
{\bf Fig. 2} The two-photon luminosity $dL/dW = L_b \times dL_{\gamma
\gamma}/dW$ in coherent $NN$ and
$pp$ collisions. The values for $L_b$ are shown in Table 1. The upper and lower
dotted curves
correspond to a different impact parameter cut-off for $pp$ collisions
for which the lower luminosity was used. From ref.[8].

\noindent
{\bf Fig. 3}
\noindent The total cross-section for the production of the standard model
Higgs via two-photon
fusion in coherent $NN$ collisions (3a) and in $pp$ collisions
at the upgrated luminosity (3b). From ref.[8].

\noindent
{\bf Fig. 4}
\noindent The two-photon production of a pair of charginos (upper curve) and a
pair of
sleptons (lower curve) as a function of the particles masses in coherent Ca-Ca
collisions
at the LHC.

\bye